\documentclass{article}

\RequirePackage[l2tabu,orthodox]{nag} 

\usepackage{arxiv}

\RequirePackage{ifxetex, ifluatex}
\newif\ifxetexorluatex   
\ifxetex
  \xetexorluatextrue
\else
  \ifluatex
    \xetexorluatextrue
  \else
    \xetexorluatexfalse
  \fi
\fi

\usepackage{amsthm,amsmath,amscd} 
\usepackage{amsfonts,amssymb}     
\usepackage{mathtools}            
\usepackage{mathtext}
\usepackage{cancel}

\ifxetexorluatex
  
\fi
\usepackage[colorlinks=true,unicode=true]{hyperref}

\usepackage{xspace} 
\usepackage{color}
\usepackage{enumitem}
\usepackage{cmap}
\usepackage{array}
\usepackage{braket}
\usepackage{epsfig}
\usepackage{float}
\usepackage{caption}
\usepackage{subcaption}
\usepackage{indentfirst}
\usepackage[normalem]{ulem}
\usepackage{hyphenat}
\usepackage{nccstretch} 
\usepackage[nice]{nicefrac}
\usepackage{stackengine}

\usepackage{cleveref} 
\creflabelformat{equation}{#2#1#3} 
\crefrangelabelformat{equation}{#3#1#4 -- #5#2#6}

\usepackage{kvsetkeys}
\makeatletter
\let\org@@cref\@cref
\renewcommand*{\@cref}[2]{%
  \edef\process@me{%
    \noexpand\org@@cref{#1}{\zap@space#2 \@empty}%
  }\process@me
}
\makeatother

\ifxetexorluatex
  \renewcommand{\mathbf}{\ensuremath{\symbf}}
  \usepackage{unicode-math}
  \usepackage{polyglossia}                        
  \setmainlanguage{english}
  \defaultfontfeatures{Ligatures=TeX}             
  \makeatletter
  \newlength\fake@f
  \newlength\fake@c
  \def\textsc#1{%
    \begingroup%
    \xdef\fake@name{\csname\curr@fontshape/\f@size\endcsname}%
    \fontsize{\fontdimen8\fake@name}{\baselineskip}\selectfont%
    \MakeUppercase{#1}%
    \endgroup%
  }
  \makeatother

\else
  \usepackage[T2A]{fontenc}           
  \usepackage[utf8]{inputenc}         
  \usepackage[main=english]{babel} 
\fi

\renewcommand{\thesubsubsection}{(\roman{subsubsection})}
\usepackage{titlesec}
\titleformat{\section}[hang]{\normalfont\large}{\bfseries\thesection.}{.5em}{\bfseries}
\titlespacing{\section}{0pt}{12pt plus 6pt}{3pt}
\titleformat{\subsection}[runin]{\normalfont}{\bfseries\thesubsection.}{.5em}{\bfseries}[.\quad]
\titleformat{\subsubsection}[runin]{\normalfont}{\bfseries{{\thesubsubsection}}}{.5em}{\bfseries}[.\quad]
\titlespacing{\subsubsection}{0pt}{0pt}{3pt}
\makeatletter
\renewcommand\l@section{\@dottedtocline{1}{0ex}{2em}}
\makeatother

\usepackage{fancyhdr}
\usepackage[format=plain,labelfont=it,font=normalsize,skip=6pt]{caption}
\DeclareCaptionLabelSeparator{dot}{.~} 
\DeclareCaptionFormat{LabRCaptC}{\hfill#1#2#3}

\usepackage{csquotes} 
\makeatletter
\usepackage[%
backend=biber, 
bibencoding=utf8, 
sorting=none, 
style=gost-numeric, 
language=autobib, 
autolang=other, 
clearlang=true, 
defernumbers=true, 
sortcites=true, 
movenames=false,  
minnames=3,  
maxnames=4,  
doi=true, 
isbn=false, 
url=false,
eprint=true,
backref=true
]{biblatex}[2016/09/17]
{\toggletrue{bbx:gostbibliography}%
}{}
\makeatother

\DefineBibliographyStrings{english}{
  backrefpage  = {Cit. on p.\adddot},
  backrefpages = {Cit. on pp.\adddot},
}

\DeclareMultiCiteCommand{\multicites}[\mkbibbrackets]{\cite}{\addsemicolon\space}

\usepackage[dvipsnames]{xcolor}
\definecolor{linkcolor}{rgb}{0.08, 0.38, 0.74}
\definecolor{citecolor}{rgb}{0.18, 0.55, 0.34}
\definecolor{urlcolor}{rgb}{0.03, 0.57, 0.82}

\hypersetup{
  linktocpage=true,           
  colorlinks,                 
  linkcolor={linkcolor},      
  citecolor={citecolor},      
  urlcolor={urlcolor},        
}
\ifluatex
  \hypersetup{
    unicode,                
  }
\fi

\usepackage[section]{placeins}
\makeatletter
\AtBeginDocument{%
  \expandafter\renewcommand\expandafter\subsection\expandafter
  {\expandafter\@fb@secFB\subsection}%
}
\makeatother
\usepackage{wrapfig}

\renewcommand{\leq}{\leqslant}

\renewcommand{\le}{\leqslant}

\emergencystretch=\maxdimen
\newcommand{\code}[1]{\texttt{#1}}

\def\phantomsph{\code{PHAN\-TOM}\xspace}

\graphicspath{{pics/}}

\makeatletter
\@ifundefined{c@bibliography_from_bbl}{
  \newcounter{bibliography_from_bbl}
  \setcounter{bibliography_from_bbl}{1}
}{}
\makeatother

\ifnum \value{bibliography_from_bbl}=0
\else
\usepackage[bblfile=phantom_com_arxiv]{biblatex-readbbl}
\fi

\title{Non-conservation of linear momentum in widely used
  hierarchical methods in gravitational gas dynamics}

\date{2024 year}

\usepackage{authblk}

\newbox{\orcid}\sbox{\orcid}{\includegraphics[scale=0.06]{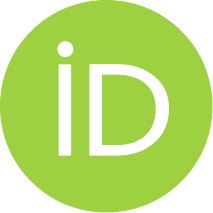}}
\author[1,2]{%
  \href{https://orcid.org/0000-0002-0564-1101}{\usebox{\orcid}\hspace{1mm}Marat~Sh.~Potashov\thanks{\texttt{marat.potashov@gmail.com}}}%
}
\author[2]{%
  \href{https://orcid.org/0000-0002-0986-4257}{\usebox{\orcid}\hspace{1mm}Andrey~V.~Yudin}%
}
\affil[1]{Keldysh Institute of Applied Mathematics, Miusskaya sq., 4, Moscow, Russia}
\affil[2]{National Research Center ``Kurchatov Institute'', Kurchatov sq., 1, Moscow, Russia}

\renewcommand{\shorttitle}{Non-conservation of linear momentum}

\hypersetup{
pdftitle={Non-conservation of linear momentum in widely used
  hierarchical methods in gravitational gas dynamics},
pdfauthor={Marat~Sh.~Potashov, Andrey~V.~Yudin},
pdfkeywords={tree code,
  fast multipole method, FMM,
  momentum conservation,
  N-body,
  smoothed particle hydrodynamics, SPH,
  PHANTOM}
}

\begin{document}
\maketitle

\begin{abstract}
  The paper considers the implementation of the fast multipole method (FMM)
    in the \phantomsph code for the calculation of forces in a self-gravitating system.
  The gravitational interaction forces are divided into short-range and long-range
    interactions depending on the value of the tree opening parameter of the hierarchical kd-tree.
  It is demonstrated that Newton's third law holds for any pair of cells
    of the kd-tree engaged in mutual interaction.
  However, for the entire system a linear momentum is not conserved.
  As a result, there is an unphysical force that causes the center of mass to migrate.
  For example, for a pair of neutron stars,
    the displacement of the system's center of mass is found to be comparable to the
    radii of the objects at times of a few tens of Keplerian revolutions.
  This displacement cannot be reduced by increasing the number of particles for values of the tree opening
    parameter greater than 0.2.
  For smaller values, the time required for the calculation is significantly longer.
\end{abstract}

\keywords
{tree code \and
  fast multipole method \and FMM \and
  momentum conservation \and
  N-body \and
  smoothed particle hydrodynamics \and SPH \and
  PHANTOM}

\section{Introduction}\label{sec:introduction}
The number of computations required to accurately calculate the forces in
  a self-gravitating system, acting between a set of \(N\) particles, grows with the
  increase of the number of particles as a function of \(\mathcal{O}(N^{2})\).
This makes calculations involving tens of thousands of particles practically impossible.
Several alternatives have been proposed to solve the problem of quadratic growth in
  computational costs for self-gravity.
These include the tree-code
  method~\cite{BarnesHut1986},
  the fast multipole method (FMM)
  \cite{GreengardRokhlin1997,
    Capuzzo-DolcettaMiocchi1998,
    bookGumerovDuraiswami2004,
    Gumerov2013},
  and hybrid methods combining elements of particle mesh
  approaches~\cite{bookHockneyEastwood2021}
  and the FMM \cite{Nitadori2014, Wang2021}.

The computational complexity of Barnes-Hut method \cite{BarnesHut1986}
  is \(\mathcal{O}(N \log N)\).
Many FMM implementations, including those presented in
  \cite{Dehnen2000, Dehnen2002, GaftonRosswog2011},
  yield a linear complexity
  \(\mathcal{O}(N)\).
Moreover, there are even such realizations of FMM~\cite{Dehnen2014}
  whose computational cost achieves \(\mathcal{O}(N^{0.87})\) operations with comparable errors.

The aforementioned methods are based on the concept of collecting particles into
  hierarchically structured groups (cells) forming a tree.
This enables to carry out the
  costly computational calculations, wherein the forces of
  gravitational attraction are taken into account at the level of the cell as a whole,
  rather than on an individual particle basis.
Different codes use various kinds of trees.
An octree, which is generated by dividing each cell into eight subcells,
  is considered within the tree-code
  \cite{BarnesHut1986, Dehnen2000, Dehnen2002}
  and \code{Octo-Tiger} \cite{MarcelloEtAl2021}.
The binary kd-tree \cite{Bentley1975} is employed by the code \code{PKDGRAV3}
  \cite{PotterEtAl2017}, which is used in cosmological simulations.
Additionally, it is utilised in codes based on the method of the smoothed particle
  hydrodynamics (SPH) such as \cite{GaftonRosswog2011} and
  \phantomsph~\cite{Price2012, PriceWursterTriccoEtal2018}.

This article discusses the implementation of the FMM in the 3D SPH magnetohydrodynamic
  code \phantomsph.
In computational astrophysics, the FMM is one of the most prevalent,
  while the \phantomsph code is successfully used in various fields of astrophysics
  \cite{2018ApJ...860L..13P,
    2019ApJ...872..163G,
    2020A&A...641A..64H,
    BlinnikovEtAl2022,
    PotashovYudin2023,
    YudinEtAl2023}.
The current implementation of FMM in \phantomsph in Cartesian coordinates
  demonstrates the complexity \(\mathcal{O}(N \log N)\).
However, as will be showed,
  this implementation does not conserve the total linear momentum of the system.
This article complements and revises an earlier paper \cite{Potashov2024}.

The article is structured in the following way.
In the section \ref{sec:phantom_grav}
  we briefly describe the \phantomsph code and its implementation of the method for
  calculating the self-gravity forces.
In the section \ref{sec:symmetric} we prove the
  fulfilment of Newton's third law for any pair of kd-tree cells in the case of mutual
  interaction.
In section \ref{sec:asymmetric} it is shown that
  the way of describing the interaction for the whole system in the \phantomsph code
  does not satisfy the third law, resulting in the generation of a non-physical force.
In the section \ref{sec:com} we demonstrate how the center of mass
  of the system migrates due to this force.
In the section \ref{sec:phantom_nonconservo} the non-conservation
  of linear momentum for the whole system in the \phantomsph calculations
  is illustrated by the example of a single neutron star (NS).
It is also shown which factors affect this
  non-conservation. In the final section \ref{sec:conclusion} we conclude that there
  is a need to change the code \phantomsph, for example, based on the works
  \cite{Dehnen2000, Dehnen2002}.

\section{Description of the self-gravity forces in \phantomsph}
\label{sec:phantom_grav}

The \phantomsph code uses SPH~--- it's a Lagrangian meshless method.
Particles in the SPH method are volumetric elements
  of the medium with unspecified shape,
  which are assigned to physical characteristics:
  coordinates, speed, mass, density, characteristic size,
  temperature, pressure and so on.
A restriction is introduced in \phantomsph:
  all particles have the same mass.
Discrete representation of the medium as a set of smoothed particles
  implies replacement of continuous characteristics~\(f(r)\)
  by piecewise-constant quantities \(f_i\).
For each particle \(i\), these values are calculated by summing the values \(f_j\)
  of neighbouring particles \(j\),
  where each summand is weighted by a special function called the smoothing kernel.
The approximation of spatial derivatives in the right-hand sides
  of the conservation laws equations in SPH
  is achieved by transferring the particle coordinate derivatives to
  the derivative of the smoothing kernels.
Using the solutions of the equations of motion, continuity, and energy, in the form of the SPH approximation,
  one can describe how the density, temperature, and pressure of matter
  described by SPH particles will change \cite{Price2012}.

Henceforth, we'll use a system of units,
  in which the unit of time is
\begin{equation}\label{eq:u_time}
  u_\mathrm{time} = \sqrt{\frac{u^3_{\mathrm{dist}}}{\mathrm{G} u_\mathrm{mass}}},
\end{equation}
  where \(u_\mathrm{dist}\) and \(u_\mathrm{mass}\)
  are the units of distance and mass
  respectively and
  \(G\) is the Newtonian constant of gravitation.

To obtain the acceleration of SPH-particles \(\boldsymbol{a}_{\mathrm{grav}}\)
  caused by the self-gravitational forces,
  we need to find the gravitational potential \(\Phi\)
  that satisfies the Poisson equation,
  which is written in our system of units as
\begin{equation}\label{eq:poisson}
  \nabla^2 \Phi=4 \pi \rho(\boldsymbol{r}),
\end{equation}
  where \(\rho\) is the density of the matter.
Then the corresponding acceleration will be
  \(\boldsymbol{a}_{\mathrm{grav}}=-\nabla \Phi\).

The elliptic type equation (\ref{eq:poisson}) implies an instant action.
Consequently, the solution must be global for the whole system.
One chosen particle is affected by all other particles, both near and far.
Such a solution is found in \phantomsph by representing
  of the total acceleration of the SPH-particle
  as the sum of the accelerations
  resulting from the short-range and long-range interactions:
\begin{equation}\label{eq:acceleration}
  \boldsymbol{a}_{\mathrm {grav}} =
    \boldsymbol{a}_{\mathrm {short}} + \boldsymbol{a}_{\mathrm {long}}.
\end{equation}
The splitting (\ref{eq:acceleration}) is dictated
  by the considerations of computational optimization.
The way of calculating the two types of acceleration is different.

To determine which interactions are considered
  to be short-range and which are to be long-range,
  the following procedure is performed.
All SPH-particles of the system
  are hierarchically grouped into cells of the kd-tree.
The algorithm recursively cuts the cells through their
  centers of mass and forms new subcells
  by splitting the longest axis into two parts
  in order to leave the cells with a more compact shape.
This allows to reduce
  errors when cutting off their multipole expansion
  \cite{GaftonRosswog2011}.
The procedure stops for cells containing not more than 10 SPH-particles.
Such cells are called \emph{leaf cells},
  and those higher up in the hierarchy are called \emph{super-cells}.

Let us consider two arbitrary not necessarily leaf cells
  \(\alpha\) and \(\beta\),
  the distance between their centers of mass is denoted by \(r\).
The particles of the first cell interact with the second cell in a short-range manner
  when one of the two criteria is satisfied.
The first of them is the tree opening criterion:
\begin{equation}\label{eq:criterion:tree_opening}
  \theta^2 < \left(\frac{s_{\beta}}{r}\right)^2,
\end{equation}
  where \({0 \leq \theta \leq 1}\) is the tree opening parameter.
Here, \(s\) is the size of the cell,
  which is equal to the minimum radius of the sphere centered on the center of mass of the cell,
  and containing all its SPH-particles.
Note also that for \({\theta = 0}\), the tree opening criterion
  (\ref{eq:criterion:tree_opening}) is always satisfied
  and
  \(\boldsymbol{a}_{\mathrm{grav}} =
    \boldsymbol{a}_{\mathrm{short}}\).
In this case, all accelerations in the system are calculated by direct summation.

The second criterion describes the possibility
  of intersecting smoothing spheres for SPH-particles of different cells:
\begin{equation}\label{eq:criterion:intersection}
  r^2 < \left[s_{\alpha} + s_{\beta} +
    \max(R_\mathrm{kern}h^{\alpha}_\mathrm{max}, R_\mathrm{kern}h^{\beta}_\mathrm{max})\right]^2.
\end{equation}
Here \(h_{\mathrm{max}}\) is the maximum smoothing length among all the particles in the cell,
  and \(R_\mathrm{kern}\) is the dimensionless cut-off radius of the smoothing kernel \(W(h)\).
A sphere of radius \(R_\mathrm{kern} h\) is a compact support of the kernel \(W(h)\).
If distances from the center of the particle are larger than \(R_\mathrm{kern}h\),
  all physical quantities described by SPH-particle are assumed to be zero.

The near acceleration \(\boldsymbol{a}_{\mathrm{short}}\)
  is calculated by direct summation over neighbouring particles
  (see \cite{PriceWursterTriccoEtal2018, PriceMonaghan2007}).
Gravitational potential for SPH-particles, smoothed by the kernel \(\phi(\epsilon)\)
  for points with \({r > R_\mathrm{kern} \epsilon}\), behaves like \(1/r\).
A function has no compact support.
If one uses only near acceleration to calculate the gravitational force
  for the whole system, then all the conservation properties are fulfilled,
  namely conservation of linear momentum, angular momentum and energy
  \cite{PriceWursterTriccoEtal2018, PriceMonaghan2007}.

The following procedure is used to calculate the long-range acceleration
  component \(\boldsymbol{a}_{\mathrm{long}}\) in \phantomsph.
The components of the gravitational acceleration
  of a particular cell \(\alpha\)
  caused by the gravitational attraction of another cell \(\beta\)
  are obtained by multipole expansion of the acceleration in powers of \(1/r\):
\begin{equation}\label{eq:acceleration:long:node}
  a_{\beta \rightarrow \alpha, i}(\boldsymbol{r}) = -\frac{M_{\beta}}{r^2} \hat{r}_i +
    \frac{1}{r^4}\left(\hat{r}_k Q_{\beta, i k}-\frac{5}{2} \hat{r}_i \hat{r}_j \hat{r}_k Q_{\beta, j k}\right).
\end{equation}
Here \(\boldsymbol{r}\) is a vector connecting the centers of mass of cells \(\alpha\) and \(\beta\),
  \(r\) is its length,
  and \(\hat{r}_i\) are the components of the corresponding unit vector,
  \(M_{\beta}\) is the total mass of the cell \(\beta\),
\begin{equation}\label{eq:qij}
  Q_{\beta, i j} = \sum \limits_{\boldsymbol{y} \in \beta} m_\mathrm{p}\left[3 y_i y_j - y^2 \delta_{i j}\right]
\end{equation}
  is the quadrupole moment of the cell \(\beta\),
  \(\boldsymbol{y}\) is the radius vector starting at the center of mass of the cell
  to some particle inside the cell,
  and \(m_\mathrm{p}\) is the mass of the particle,
  which in \phantomsph
  is assumed to be the same for all particles.

Consider some SPH-particle inside the cell \(\alpha\)
  with the radius vector \(\boldsymbol{x}\)
  starting at the center of mass of the cell \(\alpha\).
The components of the gravitational acceleration of this SPH-particle
  caused by interaction with the \(\beta\) cell are obtained by Taylor expansion
  of \(a_{\alpha \rightarrow \beta, i}\) up to the second order
  for small displacements \(\boldsymbol{x}\) from \(\boldsymbol{r}\):
\begin{equation}\label{eq:acceleration:long:particle}
  a_{\beta \rightarrow \alpha, i}(\boldsymbol{r}, \boldsymbol{x}) = a_{\beta \rightarrow \alpha, i}(\boldsymbol{r}) +
    x_j \frac{\partial a_{\beta \rightarrow \alpha, i}(\boldsymbol{r})}{\partial r_j} +
    \frac{1}{2} x_j x_k
    \frac{\partial^2 a_{\beta \rightarrow \alpha, i}(\boldsymbol{r})}{\partial r_j \partial r_k}.
\end{equation}

This approach is called the fast multipole method
  \cite{GreengardRokhlin1997}.
In \phantomsph it is implemented in Cartesian coordinates.
Using the notation of the paper \cite{Dehnen2014},
  one can state that the steps required to determine the acceleration \(\boldsymbol{a}_{\mathrm{long}}\)
  in \phantomsph are:
  step P2M (particle to multipole), consisting in calculation of \(Q_{ij}\) in accordance to equation~(\ref{eq:qij}),
  step M2L (multipole to local expansion), involving the use of equation~(\ref{eq:acceleration:long:node}),
  and finally step L2P (local expansion to particle) that includes the utilization of equation (\ref{eq:acceleration:long:particle}).

In the next section we prove the fulfillment of Newton's third law
  for any pair of cells,
  all particle accelerations in which are described by the formula
  (\ref{eq:acceleration:long:particle}).

\begin{figure}[!htb]
  \centering
  \includegraphics[width=0.9\textwidth]{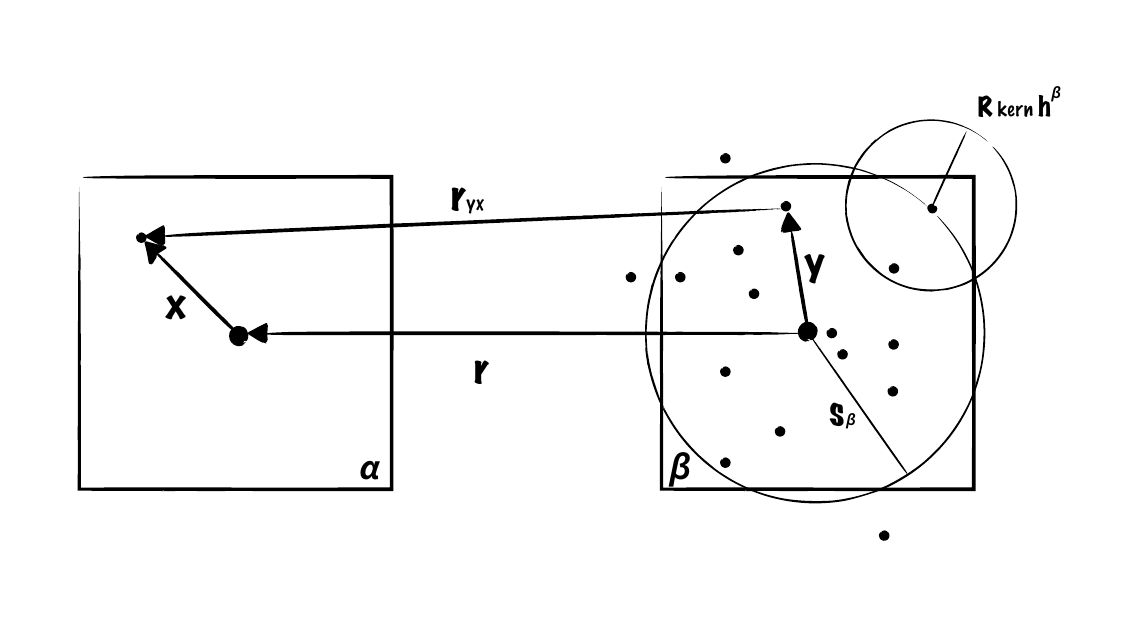}
  \caption{
    Gravitational interaction of
      SPH-particles of the \(\alpha\) cell
      with SPH-particles of the \(\beta\) kd-tree cell.
    Details in the text.
  }
  \label{fig:cells}
\end{figure}
%

\section{Symmetric interaction of a pair of kd-tree cells}
\label{sec:symmetric}

Consider a pair of kd-tree cells \(\alpha\) and \(\beta\)
  (see Figure~\ref{fig:cells}),
  which are sufficiently distant from each other
  so that their parameters do not satisfy the
  criteria~(\ref{eq:criterion:tree_opening}),~(\ref{eq:criterion:intersection}).
In this case, we will refer to such cells as \emph{well-separated}.
The minimum radius of
  the sphere located at the center of mass
  of the cell \(\beta\) containing all its
  SPH-particles, designated as \(s_{\beta}\),
  and the maximum smoothing length among all particles of the cell \(\beta\),
  designated as \(h^{\beta}_{\mathrm{max}}\), are shown in Fig.~\ref{fig:cells}.
The cells \(\alpha\) and \(\beta\) are not necessarily leaf cells.
We denote the total cell masses as \(M_{\alpha}\) and \(M_{\beta}\) correspondingly.

Let the radius vectors of SPH-particles
  inside the cells \(\alpha\) and \(\beta\)
  originating from their centers of mass
  be denoted by \(\boldsymbol{x}\) and \(\boldsymbol{y}\), respectively.
Let us also introduce a vector \(\boldsymbol{r}\) between
  the centers of mass of the cells
  \(\alpha\) and \(\beta\)
  and its length \({r = |\boldsymbol{r}|}\).
Then the distance between two arbitrary particles of different cells is
\begin{equation}\label{eq:r_xy}
  r_{yx} = |\boldsymbol{r} + \boldsymbol{x} - \boldsymbol{y}|.
\end{equation}

In order to determine the potential
  created by all particles of the cell \(\beta\)
  at the point \(\boldsymbol{x}\) of the cell \(\alpha\),
  we need to find the solution of
  the Poisson equation (\ref{eq:poisson})
  in the outer region of the \(\beta\) zone.
We will assume the potential to be equal to zero at infinity.
The solution of such a problem (see, e.g., \cite{bookBogolyubovEtAl2018})
  is defined by the expression
\begin{equation}\label{eq:phi:green_function:int}
  \Phi_{\beta \rightarrow \alpha}(\boldsymbol{x}) = \int \limits_{\beta} \rho(\boldsymbol{y}) G(\boldsymbol{r} + \boldsymbol{x}, \boldsymbol{y}) \mathrm{d}^3y,
\end{equation}
  where \(G(\boldsymbol{r} + \boldsymbol{x}, \boldsymbol{y})\)
  is the Green's function of the Laplace operator,
  which is symmetric with respect to permutation
\begin{equation}\label{eq:permutation}
  (\alpha, \boldsymbol{x}, \boldsymbol{r}) \leftrightarrows (\beta, \boldsymbol{y}, - \boldsymbol{r}).
\end{equation}
It can be shown \cite{PriceMonaghan2007}
  that for a variable smoothing length \(h(\boldsymbol{y})\)
  which depends on the density and is a function of the spatial variable,
  the expression (\ref{eq:phi:green_function:int}) in the SPH approximation converts to
\begin{equation}\label{eq:phi:green_function:sum}
  \Phi_{\beta \rightarrow \alpha}(\boldsymbol{r}, \boldsymbol{x}) =
    \sum \limits_{\boldsymbol{y} \in \beta}
    m_\mathrm{p} G(\boldsymbol{r} + \boldsymbol{x}, \boldsymbol{y}),
\end{equation}
  where
\begin{equation}\label{eq:phi:green_function:variable_h}
  G(\boldsymbol{r} + \boldsymbol{x}, \boldsymbol{y}) =
    - \frac{\phi(r_{yx}, h(\boldsymbol{x})) +
    \phi(r_{yx}, h(\boldsymbol{y}))}{2}.
\end{equation}
Here the smoothing kernel \(\phi\)
  for the potential
  is determined from the solution of the Poisson equation
\[
  W(z,h) = \frac{1}{4 \pi z^2}
    \frac{\partial}{\partial z}
    \left(z^2 \frac{\partial \phi}{\partial z}\right),
\]
  where \(z \equiv r_{yx}\).
For the standard smoothing kernels \(W(z,h)\) and from (\ref{eq:phi:green_function:variable_h}) for
\begin{equation}\label{eq:biiger_compact_support}
  r_{yx} > R_{\mathrm{kern}} h
\end{equation}
  one can obtain (see \cite{PriceWursterTriccoEtal2018, PriceMonaghan2007})
\begin{equation}\label{eq:green_function}
  G(\boldsymbol{r} + \boldsymbol{x}, \boldsymbol{y}) = - \frac{1}{r_{yx}}.
\end{equation}

The \(i\)-component of the acceleration of a SPH-particle
  at the point \(\boldsymbol{x}\)
  caused by its attraction by the entire cell \(\beta\) is
\begin{equation}\label{eq:force:particle}
  a_{\beta \rightarrow \alpha, i}(\boldsymbol{r}, \boldsymbol{x}) =
    - \frac{\partial \Phi_{\beta \rightarrow \alpha}(\boldsymbol{r}, \boldsymbol{x})}{\partial x_i} =
    - \sum \limits_{\boldsymbol{y} \in \beta} m_\mathrm{p} \frac{\partial G(\boldsymbol{r} + \boldsymbol{x}, \boldsymbol{y})}{\partial x_i}.
\end{equation}
The components of the total force acting on the cell \(\alpha\) from the side of
  cell \(\beta\)
  are obtained by summing (\ref{eq:force:particle})
  for all SPH-particles of the cell \(\alpha\):
\begin{equation}\label{eq:force:node}
  F_{\beta \rightarrow \alpha, i}(\boldsymbol{r}) =
    - \sum \limits_{\boldsymbol{x} \in \alpha,\,\boldsymbol{y} \in \beta} m_\mathrm{p}^2 \frac{\partial G(\boldsymbol{r} + \boldsymbol{x}, \boldsymbol{y})}{\partial x_i}.
\end{equation}
Due to the symmetry of the Green's function
  with respect to permutation (\ref{eq:permutation}),
  the resulting expression is symmetric:
  \(F_{\beta \rightarrow \alpha, i} = F_{\alpha \rightarrow \beta, i}\).
Since inequality (\ref{eq:biiger_compact_support}) is fulfilled for well-separated cells,
  the Green's function has a simple form (\ref{eq:green_function}).
Therefore, the gravitational interaction force (\ref{eq:force:node}) between such cells
  does not depend on the parameters of the SPH method.
This approach is consistent with the N-body formalism.

By summing (\ref{eq:force:particle}) over all cells \(\beta\),
  we get the total acceleration of the SPH-particle
  at the point \(\boldsymbol{x}\).
Even taking into account the symmetry of the Green's function
  with respect to permutation (\ref{eq:permutation}),
  it is clear that to calculate the behavior of all \(N\) SPH-particles
  in the system it is necessary \(\propto N^2\) times
  to take the derivative of it at large \(N\).
As was described in section (\ref{sec:introduction}),
  several alternative methods have been proposed
  to solve the problem of quadratic computational complexity.
Some of them demonstrate the complexity
  of \(\mathcal{O}(N^{0.87})\)
  for comparable errors.
This speed gain is largely due to
  the recursive way of traversing the tree
  and finding mutually-symmetrically interacting cells
  \cite{Dehnen2000}.
However, all these methods share a common property:
  they replace the ``particle--particle'' interactions
  with other types of interactions, such as
  ``particle--cell'' and ``cell--cell'' interactions.
To implement this, a multipole extension is used.

The Taylor expansion of Green's function (\ref{eq:green_function})
  around \(\boldsymbol{r}\) to order \(p\) reads
\begin{equation}\label{eq:green_function:series}
  G(\boldsymbol{r} + \boldsymbol{x}, \boldsymbol{y}) =
    \sum_{|n|=0}^{p} \sum_{|m|=0}^{p - |n|}
    \frac{x^n y^m}{n! m!}
    \left. {\frac{\partial^{|n|+|m|} G(\boldsymbol{r} + \boldsymbol{x}, \boldsymbol{y})}{\partial x^n \partial y^m}}
    _{\stackunder[1pt]{}{}}
    \right|_{
    \stackon[1pt]{$\scriptscriptstyle \boldsymbol{y} = 0$}{$\scriptscriptstyle \boldsymbol{x} = 0$}}
  + R_p(\boldsymbol{r} + \boldsymbol{x}, \boldsymbol{y}).
\end{equation}
Here we use multi-indices notation
  (see, e.g., \cite{Dehnen2014, bookFolland2002}),
  which is more convenient than tensor notation in this case.
The first sum corresponds to a triple summation
  for each component of \(\boldsymbol{x}\),
  where the sum of the indices \(|n| = n_{x_1} + n_{x_2} + n_{x_3} \leq p\).
The second sum corresponds to a triple summation
  for component of \(\boldsymbol{y}\),
  where the sum of the indices \(|m| = n_{y_1} + n_{y_2} + n_{y_3} \leq p - |n|\).
The residual term of the Taylor formula is \(R_p\).
Let us define the following notations:
\begin{equation}\label{eq:green_function:series:aux}
  \begin{gathered}
    x^n \equiv x_1^{n_{x_1}} x_2^{n_{x_2}} x_3^{n_{x_3}},
    \qquad
    y^m \equiv y_1^{n_{y_1}} y_2^{n_{y_2}} y_3^{n_{x_3}},
    \\
    n! \equiv n_{x_1}!n_{x_2}!n_{x_3}!,
    \qquad
    m! \equiv n_{y_1}!n_{y_2}!n_{y_3}!,
    \\
    \partial x^n \equiv \partial x_1^{n_{x_1}} \partial x_2^{n_{x_2}} \partial x_3^{n_{x_3}},
    \qquad
    \partial y^m \equiv \partial y_1^{n_{y_1}} \partial y_2^{n_{y_2}} \partial y_3^{n_{x_3}}.
  \end{gathered}
\end{equation}
The gradient of the Green's function at the point
  \(\boldsymbol{x} = \boldsymbol{y} = 0\)
  depends only on \(\boldsymbol{r}\)
\begin{equation}\label{eq:green_function:diff}
  \left. {\frac{\partial^{|n|+|m|} G(\boldsymbol{r} + \boldsymbol{x}, \boldsymbol{y})}{\partial x^n \partial y^m}}
    _{\stackunder[1pt]{}{}}
  \right|_{
  \stackon[1pt]{$\scriptscriptstyle \boldsymbol{y} = 0$}{$\scriptscriptstyle \boldsymbol{x} = 0$}} =
  (-1)^{|m|} \partial^{n+m} G(\boldsymbol{r}, 0),
\end{equation}
where we use the following shortened form:
\begin{equation}\label{eq:green_function:diff2}
  \partial^{n+m} f \equiv \frac{\partial^{|n|+|m|} f}{\partial r^{n+m}},
\end{equation}
\begin{equation}\label{eq:green_function:series:aux2}
  \begin{gathered}
    \partial r^{n+m} \equiv \partial r_1^{n_{x_1} + n_{y_1}} \partial r_2^{n_{x_2} + n_{y_2}} \partial r_3^{n_{x_3} + n_{y_3}}.
  \end{gathered}
\end{equation}

Substituting (\ref{eq:green_function:series})
  into (\ref{eq:phi:green_function:sum})
  and taking into account (\ref{eq:green_function:diff}),
  after regrouping the terms we have
\begin{equation}\label{eq:phi:series}
  \Phi_{\beta \rightarrow \alpha}(\boldsymbol{r}, \boldsymbol{x}) =
    \sum_{|n|=0}^{p} \frac{x^n}{n!}
      \sum_{|m|=0}^{p - |n|} M_m \partial^{n + m}
      G(\boldsymbol{r}, 0)
    + \sum \limits_{\boldsymbol{y} \in \beta}
    m_\mathrm{p}R_p(\boldsymbol{r} + \boldsymbol{x}, \boldsymbol{y}),
\end{equation}
  where
\begin{equation}\label{eq:phi:series:multipoles}
  M_m =
    \sum \limits_{\boldsymbol{y} \in \beta}
    m_\mathrm{p}\frac{(-1)^{|m|}}{m!}y^m
\end{equation}
  are multipole moments.
Setting \(p = 3\) in the formula (\ref{eq:phi:series})
  and neglecting the octopoles \(M_3 = 0\),
  we obtain the expressions from \cite{Dehnen2000, Dehnen2002}.

Differentiating the gravitational potential
  (\ref{eq:phi:series})
  with respect to \(x_i\)
  and taking into account (\ref{eq:force:particle}),
  we get
\begin{equation}\label{eq:force:particle:series}
  a_{\beta \rightarrow \alpha, i}(\boldsymbol{r}, \boldsymbol{x}) =
    \sum_{|n|=0}^{p - 1} \frac{x^n}{n!}
    \partial^{n}
    a_{\beta \rightarrow \alpha, i}(\boldsymbol{r}) +
    \tilde R_p(\boldsymbol{r}, \boldsymbol{x}),
\end{equation}
  where
\begin{equation}\label{eq:force:particle:series:A}
  a_{\beta \rightarrow \alpha, i}(\boldsymbol{r}) =
    - \sum_{|m|=0}^{p - 1 - |n|}
    M_m
    \partial^{m + 1} G(\boldsymbol{r}, 0)
\end{equation}
  and \(\tilde R_p(\boldsymbol{r}, \boldsymbol{x})\) is a residual term.
For \(p = 3\), the series (\ref{eq:force:particle:series})
  is similar to the acceleration expression in
  \phantomsph~(\ref{eq:acceleration:long:particle}).
However, there are differences.
In the equation (\ref{eq:acceleration:long:particle})
  the highest degree of \(1/r\) in the third term of the corresponding Hessian matrix
  is equal to 6.
In the equation (\ref{eq:force:particle:series})
  this degree derived from \(\partial^3 G(\boldsymbol{r}, 0)\) is equal to 4.

To obtain the desired degrees, we carry out the following procedure:
  we write out the series (\ref{eq:force:particle:series}) for \(p = 5\),
  discarding all terms with
  \(n = 3\), \(n = 4\), \(m = 3\), \(m = 4\),
  excluding octopoles and higher orders of \(x\) and \(y\).
The final series differs from the series
  (\ref{eq:force:particle:series}) at \(p = 3\)
  in that it has three additional terms with pairs of indices
\begin{equation}\label{eq:indexes}
  (n, m) = (1, 2); (2, 1); (2, 2),
\end{equation}
  whose sum is less than the value
  of the residual term \(\tilde R_3(\boldsymbol{r}, \boldsymbol{x})\),
  remaining within this error.
The final expressions are
\begin{equation}\label{eq:force:particle:series2}
  a_{\beta \rightarrow \alpha, i}(\boldsymbol{r}, \boldsymbol{x}) \approx
    \sum_{|n|=0}^{2} \frac{x^n}{n!}
    \partial^{n}
    a_{\beta \rightarrow \alpha, i}(\boldsymbol{r})
\end{equation}
  and
\begin{equation}\label{eq:force:particle:series:A2}
  a_{\beta \rightarrow \alpha, i}(\boldsymbol{r}) =
    \sum_{|m|=0}^{2}
    M_m
    \partial^{m + 1} G(\boldsymbol{r}, 0).
\end{equation}
They coincide with expressions
  (\ref{eq:acceleration:long:particle}) and (\ref{eq:acceleration:long:node})
  respectively.

One can give another description of the difference between the approaches
  (\ref{eq:phi:series:multipoles})--(\ref{eq:force:particle:series:A})
  and
  (\ref{eq:acceleration:long:node})--(\ref{eq:acceleration:long:particle}).
  We will call all particles of the cell \(\alpha\)
  as \emph{sinks}
  and particles of the cell \(\beta\)
  as \emph{sources}.
Then the approach
  (\ref{eq:phi:series:multipoles})--(\ref{eq:force:particle:series:A})
  involves a Taylor expansion over small displacements from \(\boldsymbol{r}\)
  of both source and sink
  so that their combined order is not greater than \({(p - 1)}\).
And the approach
  (\ref{eq:acceleration:long:node})--(\ref{eq:acceleration:long:particle})
  is a successive expansion first by small displacements from \(\boldsymbol{r}\)
  of the source to order \({(p - 1)}\)
  and only then by small displacements
  of the receiver to order \({(p - 1)}\).

Now let's go back to the expression
  (\ref{eq:green_function:series}).
It is symmetric with respect to permutation (\ref{eq:permutation})
  at every order of precision.
However, the symmetry of this expression will not be changed
  if we carry out the procedure described above
  by adding to the series (\ref{eq:green_function:series})
  at \({p = 3}\) the pairs (\ref{eq:indexes}), which are
  symmetric by indices as well.
Substitution of the Green's symmetric function
  modified in this way into (\ref{eq:force:particle})
  gives us accelerations from \phantomsph (\ref{eq:acceleration:long:particle}).
Furthermore, substituting it into (\ref{eq:force:node}),
  we will conclude that the Newton's third law
  is fulfilled for any pair of cells,
  where the acceleration of particles is described by the formula
  (\ref{eq:acceleration:long:particle}).

\section{Asymmetric interaction of kd-tree cells}
\label{sec:asymmetric}

In the previous section,
  we saw that the total force of interaction for any pair
  of cells is zero.
This implies that if only the \emph{pairs} of mutually interacting cells
  will appear in the calculation for estimating of the self-gravity forces
  for each particle at each computational time step,
  then all forces in the whole system will be compensated with machine accuracy.
At least with the accuracy of the errors resulting from the summation
  of a large amount of floating-points numbers
  \cite{LangeRump2017, Rhyne2021, HallmanIpsen2021}.

In \phantomsph, at each calculation step after updating the kd-tree
  a paired well-separated remote cell is sought each leaf cell.
It is important that such a pair is searched for \emph{only} for leaf cells,
  but not for super-cells.
As a result, in this method the direct pairs ``leaf cell \(\leftarrow\) super-cell''
  will be encountered,
  but the reverse pairs will \emph{never} be considered,
  which leads to asymmetry.
The tree opening criterion (\ref{eq:criterion:tree_opening})
  is also asymmetric in \phantomsph because it contains parameters of one cell only.
An example of symmetric criteria is considered in~\cite{Dehnen2000, LoiseauEtAl2020}.

\begin{figure}[!htb]
  \centering
  \includegraphics[width=0.9\textwidth]{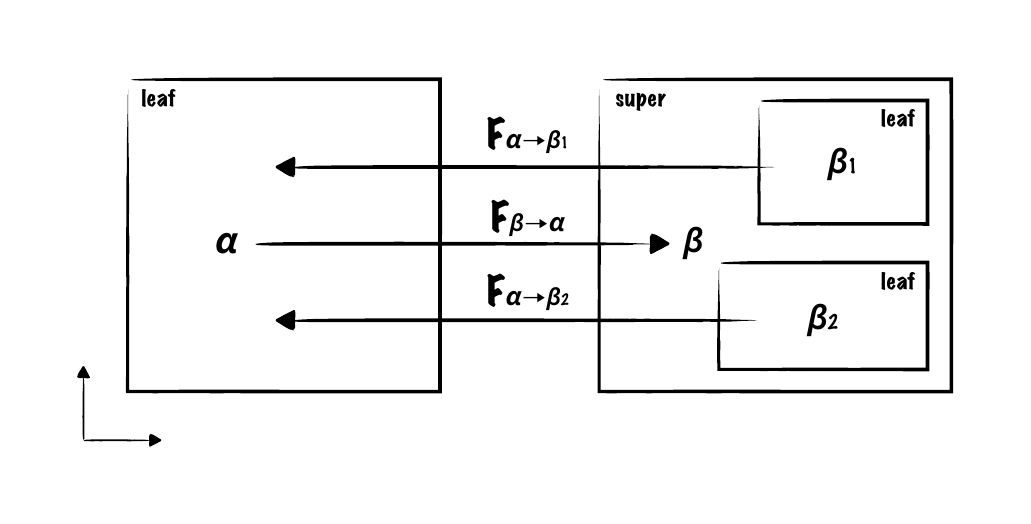}
  \caption{
    Non-symmetric gravitational interaction between SPH-particles
      of the leaf cell \(\alpha\)
      and SPH-particles of the supercell \(\beta\) of the kd-tree.
    Details in the text.
  }
  \label{fig:asymmetry}
\end{figure}

A model of such a situation is illustrated in Fig.~\ref{fig:asymmetry}.
A leaf cell \(\alpha\) with mass \(M_{\alpha}\)
  is attracted by the super-cell \(\beta\) with mass \(M_{\beta}\)
  with a force \(\boldsymbol{F}_{\beta\rightarrow\alpha}\).
Let the centers of masses of the cells \(\alpha\) and \(\beta\)
  be at the distance \(r\).
The super-cell consists of two leaf cells \(\beta_1\) and \(\beta_2\),
  which attract \(\alpha\) with the forces
  \(\boldsymbol{F}_{\alpha \rightarrow \beta_1}\) and \(\boldsymbol{F}_{\alpha \rightarrow \beta_2}\) respectively.
The masses of these cells \(M_{\beta_1}\) and \(M_{\beta_2}\) add up to \(M_{\beta}\).
Let's assume that \(M_{\beta_2} > M_{\beta_1}\).
Let the radius vector of the center of mass of the cell \(\beta_1\)
  originating from the center of mass of the cell \(\beta\) be \(\boldsymbol{x}\).
Then the radius vector of the center of mass of the cell \(\beta_2\)
  originating from the center of mass of the cell \(\beta\)
  is \(- \boldsymbol{x} M_{\beta_1}/M_{\beta_2}\),
  by the definition of the center of mass \(\beta\).

The \(i\)-component of the total force
  acting on the center of mass of the whole system is
\begin{align}\label{eq:force:tree}
  \begin{split}
    F_i = F_{\beta \rightarrow \alpha, i} +&
      F_{\alpha \rightarrow \beta_1, i} +
      F_{\alpha \rightarrow \beta_2, i} =
      \\
      &\sum \limits_{\boldsymbol{x} \in \alpha} m_\mathrm{p} a_{\beta \rightarrow \alpha, i} +
      \sum \limits_{\boldsymbol{x} \in \beta_1} m_\mathrm{p} a_{\alpha \rightarrow \beta_1, i} +
      \sum \limits_{\boldsymbol{x} \in \beta_2} m_\mathrm{p} a_{\alpha \rightarrow \beta_2, i},
  \end{split}
\end{align}
  where the expressions for accelerations are taken from
  (\ref{eq:acceleration:long:particle})
  or, equivalently, from (\ref{eq:force:particle:series2}).
The summation is carried out over the internal particles of each cell.
All cell mass centers lie in the same plane \(P\)
  (Fig.~\ref{fig:asymmetry} plane).
Without limiting generality,
  let us arrange the coordinate system
  so that one axis passes through the centers of masses of the cells \(\alpha\) and \(\beta\),
  the second axis lies in the plane \(P\),
  and the third axis is perpendicular to \(P\).

Expanding \(F_i\) in a Taylor series by small displacement \(\boldsymbol{x}\)
  up to third order and writing out the leading terms, we get
\begin{equation}\label{eq:force:tree:series}
  \begin{gathered}
    F_1 \approx \frac{2}{r^5}
      M_{\alpha} M_{\beta} \frac{M_{\beta_1}}{M_{\beta_2}}
      \left(1 - \frac{M_{\beta_1}}{M_{\beta_2}}\right)
      (2 x_1^2 - 3 x_2^2) x_1,
    \\
    F_2 \approx \frac{3}{2 r^5}
      M_{\alpha} M_{\beta} \frac{M_{\beta_1}}{M_{\beta_2}}
      \left(1 - \frac{M_{\beta_1}}{M_{\beta_2}}\right)
      (x_2^2 - 4 x_1^2) x_2.
  \end{gathered}
\end{equation}
The force component \(F_3\) is zero with this accuracy.
It is important to note that the expressions (\ref{eq:force:tree:series})
  do not involve the quadrupole moments of the cells,
  which means that the location of SPH-particles
  within each cell does not affect the force.

Maximizing the norm \(F\) of the vector \(\boldsymbol{F}\) for different \(\boldsymbol{x}\),
  we obtain its following upper bound
\begin{equation}\label{eq:force:tree:max}
  F \approx \frac{4}{r^5} l^3
    M_{\alpha} M_{\beta} \frac{M_{\beta_1}}{M_{\beta_2}}
        \left(1 - \frac{M_{\beta_1}}{M_{\beta_2}}\right),
\end{equation}
  where \(l = |\boldsymbol{x}|\) is the distance
  from the center of mass of the cell \(\beta\)
  to the center of mass of the most distant cell.
The resulting non-physical uncompensated force acts
  on each of the SPH-particles of the pair \(\alpha, \beta\).
We will call such a pair of cells \emph{asymmetric}.

Let us repeat this reasoning for the case if there are
  more than two leaf cells inside a supercell \(\beta\),
  and instead of a leaf cell \(\alpha\), there can be a supercell
  with child leaf cells \(\alpha_1, \alpha_2, \dots\)\;.
Assuming that cells with indices \(\alpha_1\), \(\beta_1\)
  have the smallest masses,
  we get a generalization of the expression (\ref{eq:force:tree:max})
\begin{align}\label{eq:force:tree:max:full}
  \begin{split}
    F \approx 4
      \sum \limits_{\alpha,\beta} \frac{M_{\alpha} M_{\beta}}{r_{\alpha \beta}^5} \Biggl[
        \frac{M_{\alpha_1}l_{\alpha_1}^3}{M_{\alpha} - M_{\alpha_1}}
          \Bigl(1 -& \frac{M_{\alpha_1}}{M_{\alpha} - M_{\alpha_1}}\Bigr)+
        \\
        +&
        \frac{M_{\beta_1}l_{\beta_1}^3}{M_{\beta} - M_{\beta_1}}
          \Bigl(1 - \frac{M_{\beta_1}}{M_{\beta} - M_{\beta_1}}\Bigr)
      \Biggr].
  \end{split}
\end{align}
Here \(l_{\alpha_1}\) and \(l_{\beta_1}\) are the distances from
  the centers of mass of \(\alpha\) and \(\beta\)
  to the centers of mass of the lightest distant cells, respectively.
Summation is performed over all possible asymmetric pairs \(\alpha\), \(\beta\),
  the distances between which are \(r_{\alpha \beta}\).
If a cell \(\alpha\) in some summand is a leaf cell,
  the first summand in square brackets in (\ref{eq:force:tree:max:full})
  is assumed to be zero, indicating that its size is also zero.
It was not implied here that all cells necessarily lie in the same plane.

At some computational step there can be both asymmetric and symmetric pairs of cells.
At the next step of the computation,
  when the kd-tree is updated, the number of pairs may change.
For example, there may be a new asymmetric pair of cells
  oriented differently in space,
  or several pairs.
The center of mass of the whole system (the whole kd-tree)
  experiences a ``kick'' \(F\) at each step in different directions.
It can be said that an uncompensated random force
  will act on the system as a whole, causing it to shift from its initial position.
The system ceases to be conservative,
  and we lose one of the main advantages of the SPH approach
  compared to the mesh method.

It is important to note that if in (\ref{eq:force:tree})
  instead of (\ref{eq:acceleration:long:particle})
  we used the accelerations from (\ref{eq:force:particle:series}) at \(p = 3\),
  the resulting force would also be \emph{non-zero}.
Discarding terms with \(1/r^5\) and above
  in the formula (\ref{eq:force:tree}) does not solve
  the problem of the additional non-physical force.
In the next section we will consider
  how the center of mass of the system
  will migrate due to its action.

\section{Displacement of the system's center of mass}
\label{sec:com}

Now let's see how far the system
  as a whole shifts when it periodically receives a ``kick''
  by a force described by an isotropic random variable \(f_i\),
  whose modulus is uniformly distributed from zero to \(F\).
The time-average impact of such a force
  is \({\langle f_i \rangle = 0}\),
  and the RMS impact
  is \({\langle f_i^2 \rangle = F^2/3}\).
Assume that the calculated time steps are equal for simplicity.
Then, for each \(\Delta t\),
  the system receives a momentum \(f_i\Delta t\),
  or, similarly, it receives an additional velocity
  \(\Delta v_i = f_i\Delta t/M\),
  where \(M\) is the mass of the system and
\begin{equation}\label{eq:vrms2}
  \langle\Delta v^2_{i}\rangle =
    \frac{F^2}{3 M} \Delta t^2.
\end{equation}

We will consider one velocity component, e.g. \(x\)-component.
The magnitudes of the velocity ``kicks''
  at successive times
  are \(\Delta v_1\), \(\Delta v_2\), and so on.
The velocity on the interval from \(k\) to \(k+1\) is
\[
  v_{k,k+1} = \sum\limits_{i=1}^{k} \Delta v_{i},
\]
  and the displacement is \(S_{k,k+1} = v_{k,k+1}\Delta t\).
The displacement on the interval from one to \(k+1\) is
\[
  S_{1,k+1} = \sum\limits_{i=1}^{k} S_{i,i+1} =
    \Delta t \sum\limits_{i=1}^{k} \Delta v_{i}(k-i+1).
\]
The average displacement \(\langle S_i \rangle\) is zero
  because \(\langle \Delta v_i \rangle = 0\),
  but the average square of the displacement is not zero
\[
  \langle S^2_{1,k+1} \rangle =
    \Delta t^2 \sum\limits_{i=1}^{k} \langle\Delta v^2_{i}\rangle(k-i+1)^2,
\]
  where we take into account that \({\langle \Delta v_i \Delta v_j \rangle = 0}\)
  at \({i \neq j}\).
The value \(\langle \Delta v^2_i \rangle\) can be considered independent on~\(i\).
Then
\[
  \langle S^2_{1,k+1} \rangle = \Delta t^2 \langle \Delta v^2_{i} \rangle
    \sum\limits_{i=1}^{k} i^2.
\]
The sum here is \(k(k+1)(2k+1)/6 \simeq k^3/3\) for \(k \gg 1\)
  and \(k = T/\Delta t\),
  where \(T\) is the observation time of the system.
Finally, for the total displacement
\[
  \langle r^2 \rangle =
    \langle S^2_x \rangle + \langle S^2_y \rangle + \langle S^2_z \rangle,
\]
  we obtain
\begin{equation}\label{eq:rrms}
  \sqrt{\langle r^2 \rangle} =
    \frac{F}{M} \sqrt{\frac{\Delta t}{3}} T^{\frac{3}{2}}.
\end{equation}
The amount of momentum transferred is random.
  This means that the velocities are the sum of many ``jumps'' in momentum space.
We can state that the action of a chaotic force
  on a system leads to random walks in momentum space.
Hence
\[
  \langle v^2_{k,k+1} \rangle = k \langle \Delta v^2_i \rangle,
\]
  where \(\langle v^2_{k,k+1} \rangle\) is the average square of the speed
  that the system has in the interval from \(k\) to \(k+1\).
If we rewrite the latter expression as a function of time
  and take into account (\ref{eq:vrms2}), we get
\begin{equation}\label{eq:vrms}
  \sqrt{\langle v^2 \rangle} = \frac{F}{M} \sqrt{\frac{\Delta t}{3}} T^{\frac{1}{2}}.
\end{equation}
This shows that the velocity increases infinitely with time.
Using (\ref{eq:vrms}) we can also define the standard deviation
  of the random variable of the linear momentum of the system
\begin{equation}\label{eq:prms}
  \sqrt{\langle p^2 \rangle} = M\sqrt{\langle v^2 \rangle}.
\end{equation}

A non-diffusive random walk
  law \({|r(T)| \propto T^{3/2}}\)
  follows from equation (\ref{eq:rrms})
  (in the case of diffusion it would be \({|r(T)| \propto T^{1/2}}\)).
This deviation arises because the system ``remembers''
  the prehistory of its motion,
  accumulating momentum
  while violating the condition of independence of successive steps.
The displacement of the system's center of mass
  due to a small chaotic change in its velocity
  for each counting step is called
  a random memory walk \cite{bookPopov2016}.

Now that we know what the force of the ``kick'' depends on,
  let's find out what affects its frequency.
In \phantomsph, at the end of each numerical step,
  the \(\Delta t\) of the next step is determined
  as the minimum of all kinds of time step constraints on all SPH-particles.
Usually, the strictest constraint during
  the calculation follows from
  the Courant--Friedrichs--Lewy
  condition \cite{PriceWursterTriccoEtal2018}.
Using this condition, we obtain
\begin{equation}\label{eq:dt}
  \Delta t = C_{\mathrm{cour}} \min \limits_\mathrm{p}
    \frac{h_\mathrm{p}}{\max \limits_n v_{\mathrm{sig}, \mathrm{p} \mathrm{n}}},
\end{equation}
where \({C_{\mathrm{cour}} = 0.3}\) by default \cite{LattanzioEtAl1986},
\begin{equation}\label{eq:h_p}
  h_\mathrm{p} = h_{\mathrm{fact}}\left(\frac{m_\mathrm{p}}{\rho_\mathrm{p}}\right)^{\frac{1}{3}}
\end{equation}
  is the smoothing length,
  \(h_{\mathrm{fact}}\)
  is a numerical parameter close to unity and
\begin{equation}\label{eq:v_sig}
  v_{\mathrm{sig}, \mathrm{p} n} =
    \alpha_\mathrm{p}^{\mathrm{AV}} c_{\mathrm{s}, \mathrm{p}}
    + \beta^{\mathrm{AV}}\left|v_{\mathrm{p} \mathrm{n}} \cdot \hat{r}_{\mathrm{p} \mathrm{n}}\right|
\end{equation}
  is the maximum signal speed,
  which is a multiplier in the
  artificial viscosity tensor \cite{PriceWursterTriccoEtal2018}.
Here \(\alpha_\mathrm{p}^{\mathrm{AV}}\), \(\beta^{\mathrm{AV}}\)
  are dimensionless constants of order of unity,
  the exact values of which are not important now,
\[
  v_{\mathrm{p} \mathrm{n}} \equiv v_\mathrm{p}-v_\mathrm{n},
  \quad
  \hat{r}_{\mathrm{p} \mathrm{n}} \equiv\left(r_\mathrm{p}-r_\mathrm{n}\right)/\left|r_\mathrm{p}-r_\mathrm{n}\right|,
\]
  and the speed of sound is
\begin{equation}\label{eq:c_s}
  c_{\mathrm{s,\mathrm{p}}}=\sqrt{\frac{\gamma P_\mathrm{p}}{\rho_\mathrm{p}}},
\end{equation}
where \(P_\mathrm{p}\) is pressure, and \(\rho_\mathrm{p}\) is a density.
The artificial viscosity for
  the SPH method contains both
  a linear velocity term at the multiplier \(\alpha_\mathrm{p}^{\mathrm{AV}}\),
  analogous to shear and bulk viscosity,
  and a second-order \mbox{Von Neumann}-Richtmyer-like term
  \cite{VonNeumannRichtmyer1950} that prevents particle interpenetration.
Let us restrict ourselves to the case
  of the polytropic equation of state of matter
\begin{equation}\label{eq:eos}
  P_\mathrm{p} = K \rho_\mathrm{p}^\gamma.
\end{equation}
Combining
  (\ref{eq:h_p})--(\ref{eq:eos})
  and substituting into (\ref{eq:dt})
  for the case of hydrostatic equilibrium
  (when \({v_{\mathrm{p} \mathrm{n}} = 0}\)),
  we obtain
\begin{equation}\label{eq:dt:propto}
  \Delta t \propto m_\mathrm{p}^{\frac{1}{3}} \rho^{\frac{(1-\gamma)}{2} - \frac{1}{3}}.
\end{equation}

In the next section we will explicitly demonstrate
  the displacement of the system's center of mass
  in the \phantomsph calculations.

\section{Non-conservation of the total
  linear momentum of the system in \phantomsph}
\label{sec:phantom_nonconservo}

To illustrate the non-conservation of the total momentum,
  we will consider the model of a single NS.
To do this, in \phantomsph
  we will ``assemble'' a ball out of \(N\) SPH-particles
  with radius \(R = 10\)~km and mass \(M = 1M_\odot\),
  with a density profile satisfying the equation of state polytropes with \({n = 1}\)
  (adiabatic index of the matter \({\gamma = 2}\)).
This simple polytrope well describes the equation of state of NS
  in the intermediate mass range \(1M_\odot \le M \le 2M_\odot\)
  \cite{GreifEtAl2020}.
All SPH-particle masses assumed to be the same and equal to
\begin{equation}\label{eq:m_p}
  m_\mathrm{p} \equiv M/N.
\end{equation}
The particles are initially arranged uniformly and isotropically
  in space and then distributed in accordance
  with the requirements of the given density profile
  and radius with preservation
  of the relative arrangement along the mass coordinate
  (stretch mapping, \cite{PriceWursterTriccoEtal2018}).
The initial velocities of all particles are zero.
The constructed star is already close to equilibrium.
The diameter of a star in equilibrium
  is of the order of the critical Jeans wavelength.
This means that it is impossible to place disturbances
  inside the star that would increase
  and it is always stable with respect
  to fragmentation into many small parts
  \cite{bookZeldovichNovikov1967}.
The goal is to bring the NS to a state of \emph{complete} equilibrium.
This process is called relaxation in \phantomsph.

The relaxation simulation starts from the aforesaid initial model
  and the complete system of dynamic equations of motion
  of self-gravitating particles is solved in the SPH approximation.
This system is described by equations~\eqref{eq:phi:series} and \eqref{eq:phi:series:multipoles}
  and their SPH approximation is given by equations~\eqref{eq:vrms2} and~\eqref{eq:rrms}
  in \cite{PriceWursterTriccoEtal2018}.

In the following calculations the cubic spline kernel \(M_4\) is used
  \multicites[eq.~17~in][]{PriceWursterTriccoEtal2018}{Schoenberg1946}
  and the compact support of the function implies that \(R_\mathrm{kern} = 2\).
The choice of \(h_\mathrm{fact} = 1.2\) for the \(M_4\) cubic spline kernel
  is based on the fact that it is slightly less than the maximum neighbour number
  that can be used while remaining stable to the pairing instability
  \cite{Price2012, PriceWursterTriccoEtal2018, Rosswog2015}.
The mean neighbour number
  can be estimated as
  \(\bar{N}_{\mathrm{neigh}} = \frac{4}{3} \pi\left(R_{\mathrm{kern}} h_{\mathrm{fact}}\right)^3\),
  which for \(M_4\) kernel corresponds to 57.9 neighbours
  for particles with a uniform density distribution.
The actual number of neighbours of SPH-particle
  occurring in calculations for the case of NS
  with \(N = 5000\) is \(\sim 65^{+15}_{-13}\)
  and with \(N = 150000\) is \(\sim 63^{+10}_{-11}\).
As one can see, these values are in good agreement with
  the estimate \(\bar{N}_{\mathrm{neigh}}\).

The average smoothing length
  is \(h_\mathrm{p} \approx 0.97\)~km for \(N = 5000\).
It follows from equations (\ref{eq:h_p}) and (\ref{eq:m_p})
  that \(h_\mathrm{p}\) decreases with increasing \(N\) as \(N^{-1/3}\).
Therefore, for the full range of \(N\) considered below,
  the values of \(h_\mathrm{p}\) are significantly smaller than \(R\),
  which results in the existence of pairs of well-separated cells.

The calculation results for the
  cases with different numbers of SPH-particles
  are presented below.
The distance unit is \({u_\mathrm{dist} = 1}\)~km,
  the mass unit is \({u_\mathrm{mass} = 1M_\odot}\),
  the time unit in accordance with (\ref{eq:u_time})
  is \({u_\mathrm{time} = 2. 75 \cdot 10^{-6}}\)~s,
  and the tree opening parameter \(\theta\)
  from (\ref{eq:criterion:tree_opening}) is \(0.5\).
We take the duration of the simulation from the following considerations:
  a close binary system of identical NSs
  with the above parameters makes dozens of orbital revolutions
  in a time \(\sim 4 \cdot 10^4\;[u_\mathrm{time}]\).

\begin{figure}[!htb]
  \centering
  \includegraphics[width=0.8\textwidth]{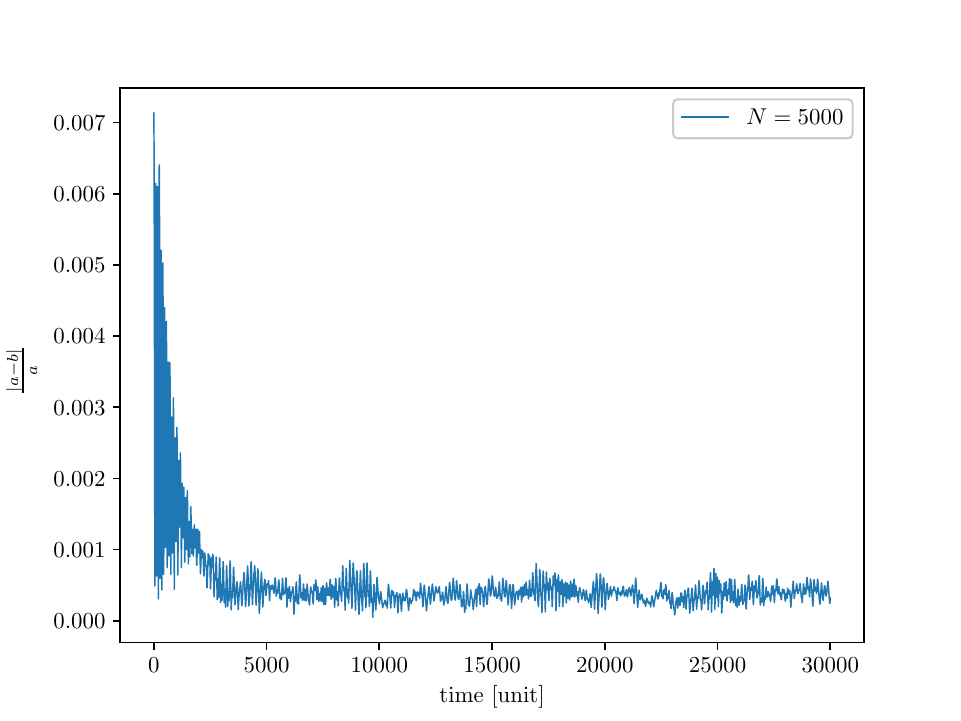}
  \caption{
    Flattening of the neutron star ellipsoid,
      calculated at SPH-particle number \(N = 5000\) and \({\theta = 0.5}\).
  }
  \label{fig:flat}
\end{figure}

The flattening \((a - b)/a\) is demonstrated at Fig.~\ref{fig:flat},
  where \(a\) and \(b\) are the semi-axes of the star's ellipsoid.
Here, the NS quickly reaches an equilibrium state
  at \(t \approx 5000\,[u_\mathrm{time}]\)
  and until the end of the simulation the star moves as a whole slightly pulsating.

\begin{figure}[!htb]
  \centering
  \includegraphics[width=0.8\textwidth]{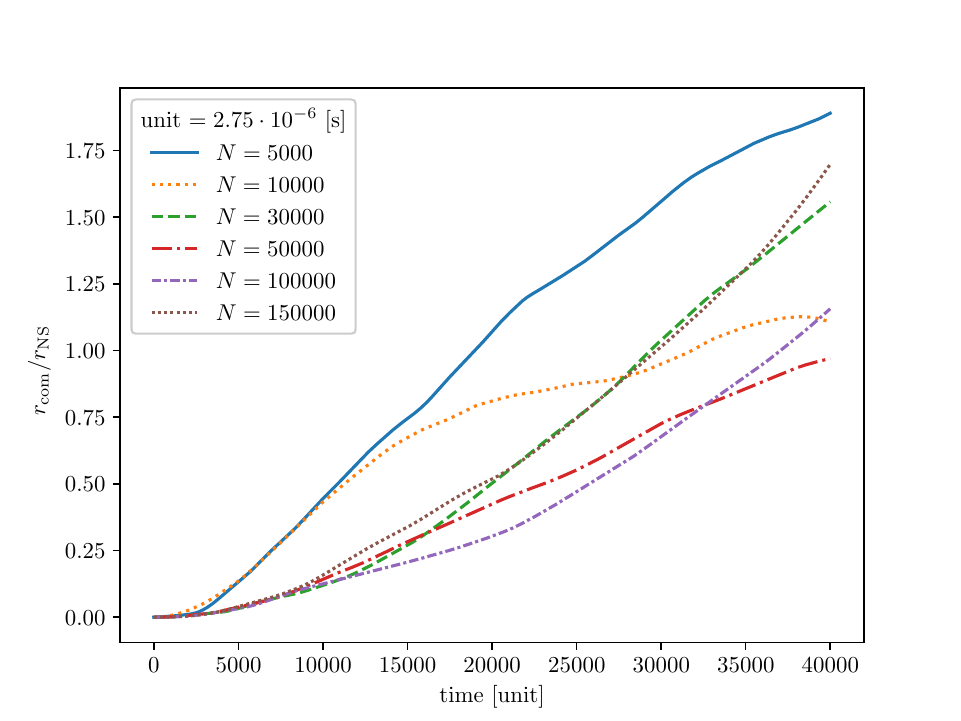}
  \caption{
    The displacement of the center of mass
      of a single neutron star
      normalized by its radius,
      calculated for six different numbers of SPH-particles at \({\theta = 0.5}\).
  }
  \label{fig:rcom}
\end{figure}

\begin{figure}[!htb]
  \centering
  \includegraphics[width=0.8\textwidth]{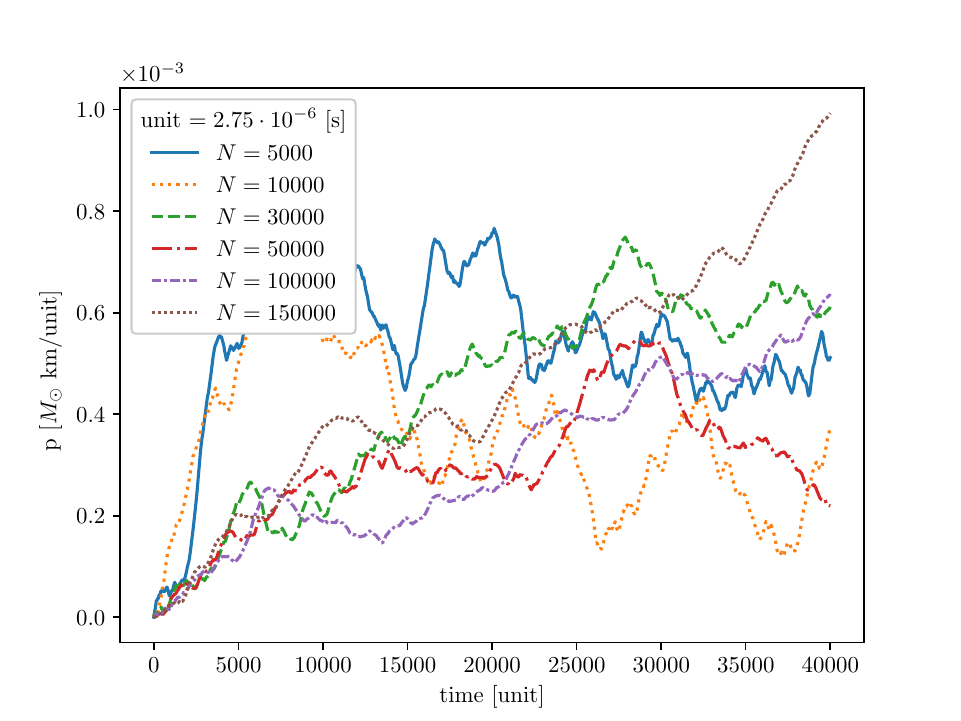}
  \caption{
    The linear momentum of a single neutron star,
      calculated for six different numbers of SPH-particles at \({\theta = 0.5}\).
  }
  \label{fig:pcom}
\end{figure}

\begin{figure}[!htb]
  \centering
  \includegraphics[width=0.8\textwidth]{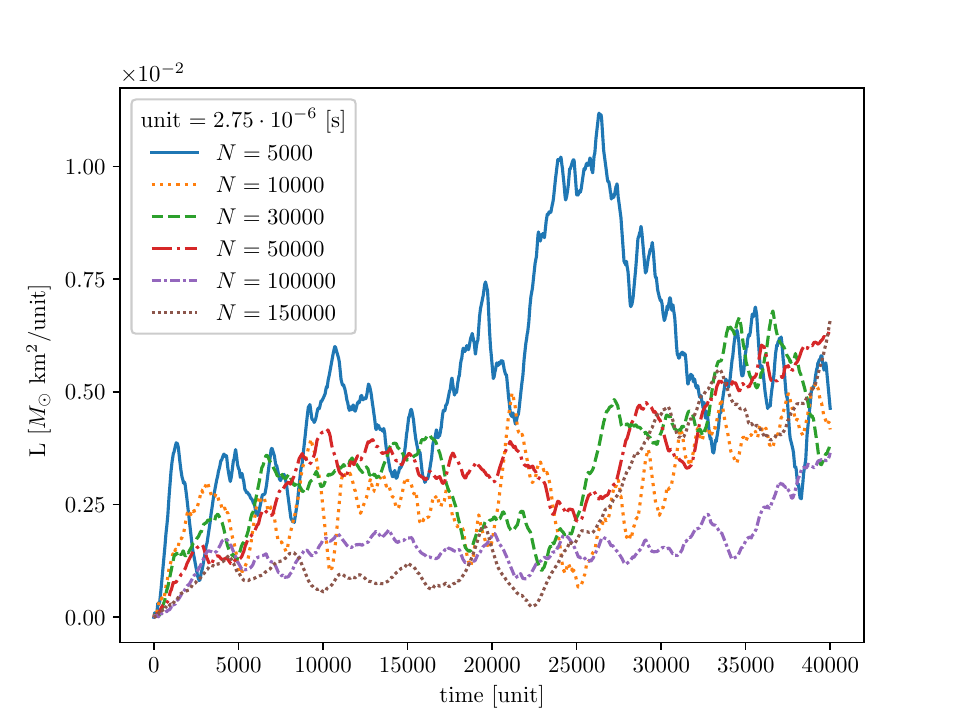}
  \caption{
    The angular momentum of a single neutron star,
      calculated for six different numbers of\\ SPH-particles at \({\theta = 0.5}\).
  }
  \label{fig:lcom}
\end{figure}

{The behavior of the random variables over time is presented
  at Fig.~\ref{fig:rcom} and \ref{fig:pcom}:
  \(r_\mathrm{com}/R\) is the deviation
  of the center of mass of the NS from the initial position,
  normalized to its radius,
  and \(p\) is the linear momentum of the NS
  for six different numbers of SPH-particles at \({\theta = 0.5}\).
The standard deviation of the random variable \(r_\mathrm{com}\)
  is defined in (\ref{eq:rrms}),
  and the standard deviation of the momentum is taken from (\ref{eq:prms}).
It can be seen (Fig.~\ref{fig:rcom}) that the NS is shifted
  from its initial position by a distance comparable to its size!
Neither the linear momentum of the system (Fig.~\ref{fig:pcom})
  nor the angular momentum are conserved (Fig.~\ref{fig:lcom}).
The latter means that the matter inside the NS has a non-zero angular velocity.
  However, the study of this observation is beyond the scope of the present article.}

\begin{figure}[!htb]
  \centering
  \includegraphics[width=0.8\textwidth]{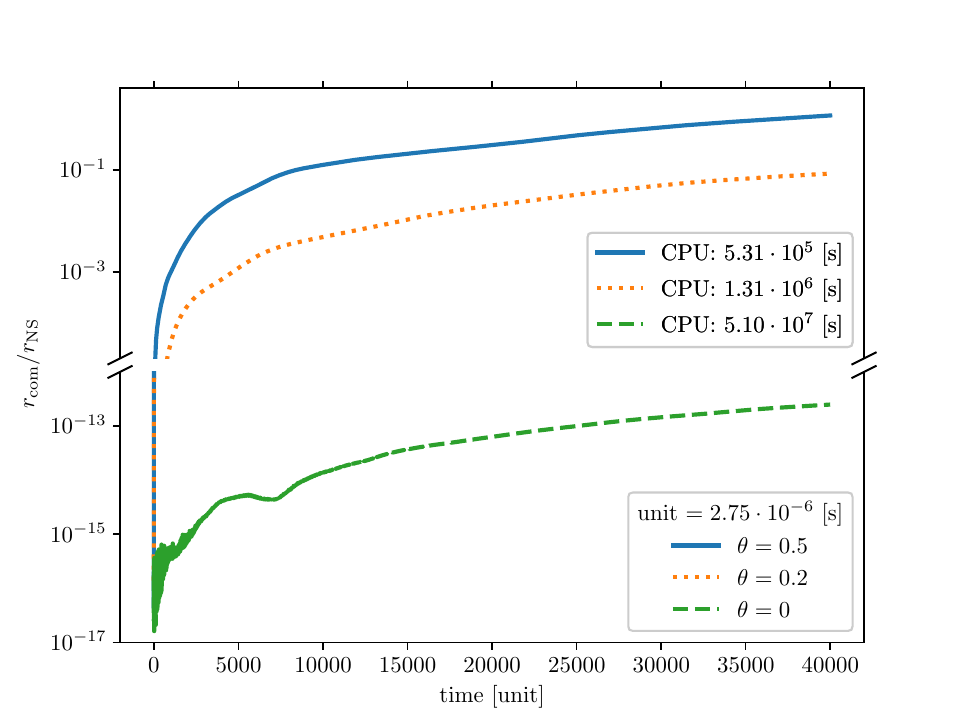}
  \caption{
    The displacement of the center of mass
      of a single neutron star
      normalized by its radius,
      calculated for three different values of \(\theta\)
      with the number of particles \(N = 100000\).
    CPU time consumption
      on AMD Ryzen 7 2700X Eight-Core Processor machine
      are presented for all three cases of value \(\theta\).
  }
  \label{fig:rcom_log_theta0-0_5}
\end{figure}

If we perform the same calculation
  with the parameter \({\theta = 0}\)
  for the tree opening criterion (\ref{eq:criterion:tree_opening}),
  when all gravitational forces in the system are calculated by direct summation,
  then there will be no such a displacement
  of the center of mass of the system
  (see Fig.~\ref{fig:rcom_log_theta0-0_5}).
This proves that the main influence on the effect of non-conservativity
  is exerted by the gravitational forces of long-range interaction
  and the displacement is not connected with SPH properties of the system.

Let us estimate which parameters affect the magnitude
  of the total linear momentum
  of the star using the results of the section \ref{sec:asymmetric}.
For estimation, we assume that all leaf cells have the same mass
\begin{equation}\label{eq:m_l}
  M_l = 10\,m_\mathrm{p}.
\end{equation}
Although in practice there are cells in which the number of SPH-particles is less than 10,
  the vast majority contains just the maximum allowable number
  of particles when constructing a kd-tree.
Then, let's introduce the numbers of leaf cells in super-cells \(\alpha\) and \(\beta\)
  are \(N_{\alpha} = M_{\alpha}/M_l\) and \(N_{\beta} = M_{\beta}/M_l\) respectively.
Finally, remembering that \(l_{\alpha_1}\), \(l_{\beta_1}\)
  are the distances to the most distant leaf cells,
  we will assume that they are also the characteristic sizes
  \(s_{\alpha_1}\) and \(s_{\beta_1}\) of their parent cells, respectively.
If \(\alpha\) is a leaf cell in some term,
  then we will assume that its size is zero.
Then (\ref{eq:force:tree:max:full}) will have the following form
\begin{align}\label{eq:force:tree:approx}
  \begin{split}
    F \approx 4 M_l^2
      \sum \limits_{\alpha,\beta} \frac{N_{\alpha} N_{\beta}}{r_{\alpha \beta}^5} \Biggl[
        s_{\alpha}^3\frac{(N_{\alpha} - 2)}{(N_{\alpha} - 1)^2} +
        s_{\beta}^3\frac{(N_{\beta} - 2)}{(N_{\beta} - 1)^2}
      \Biggr].
  \end{split}
\end{align}
When the total number of particles \(N\)
  changes in any super-cell,
  the number of its leaf cells will change
  as a piecewise-constant function
  taking values close to the function proportional to \(N/10\).
This means that \(N_{\alpha}\) and \(N_{\beta}\)
  are close to the linear function of \(N\).
Further, since every pair of cells in the sum (\ref{eq:force:tree:approx})
  by construction does not satisfy the tree opening criterion (\ref{eq:criterion:tree_opening}),
  then \((s_{\alpha}/r_{\alpha \beta})^3\)
  and \((s_{\beta}/r_{\alpha \beta})^3\)
  are majorized by the parameter \(\theta^{3/2}\).
It follows from~(\ref{eq:m_p}) and (\ref{eq:m_l})
  that \(M_l \propto M/N\),
  and any distance between cells \(r_{\alpha \beta} \propto R\).
The number of pairs of asymmetric cells in the sum (\ref{eq:force:tree:approx})
  varies proportionally to the function \(N^{\sigma+1}\),
  where \({0 \leq \sigma \leq 1}\),
  and depends on the value of the parameter \(\theta\),
  which determines the number of interaction participants.
Summarizing the foregoing,
  we get that
\begin{equation}\label{eq:force:tree:propto}
  F \propto (M/R)^2 N^\sigma.
\end{equation}

Assuming that the average density of NS is
\begin{equation}\label{eq:rho_mean}
  \rho = M \left(\frac{4}{3}\pi R^3\right)^{-1},
\end{equation}
  it is possible to obtain from (\ref{eq:m_p}) and (\ref{eq:dt:propto}) that
\begin{equation}\label{eq:dt:propto2}
  \Delta t \propto
    M^{\frac{(1-\gamma)}{2}} R^{1-\frac{3(1-\gamma)}{2}} N^{-\frac{1}{3}}.
\end{equation}
Finally, substituting
  (\ref{eq:force:tree:propto}) and (\ref{eq:dt:propto2})
  into (\ref{eq:rrms}), we get
\begin{equation}\label{eq:rrms:propto}
  \sqrt{\langle r^2 \rangle} \propto
    M^{\frac{(5-\gamma)}{4}} R^{\frac{3(\gamma - 3)}{4}}
    N^{\sigma - \frac{1}{6}} T^{\frac{3}{2}}.
\end{equation}
The obtained formula (\ref{eq:rrms:propto})
  is also true for \emph{any} ball of SPH-particles.
Let us write out three special cases:
  for~NS intermediate mass (\(\gamma = 2\))
\begin{equation}\label{eq:rrms:propto:ns}
  \sqrt{\langle r^2 \rangle} \propto
    \left(\frac{M}{R}\right)^{\frac{3}{4}} N^{\sigma - \frac{1}{6}}
    T^{\frac{3}{2}},
\end{equation}
  for a white dwarf (WD) (\({\gamma = 5/3}\))
\begin{equation}\label{eq:rrms:propto:wd}
  \sqrt{\langle r^2 \rangle} \propto
    \frac{M^{\frac{5}{6}}}{R} N^{\sigma - \frac{1}{6}}
    T^{\frac{3}{2}},
\end{equation}
  and for relativistic matter (Chandrasekhar WD or a large hot star) (\(\gamma = 4/3\))
\begin{equation}\label{eq:rrms:propto:starrel}
  \sqrt{\langle r^2 \rangle} \propto
    M^{\frac{11}{12}} R^{-\frac{5}{4}} N^{\sigma - \frac{1}{6}}
    T^{\frac{3}{2}}.
\end{equation}
It is clear from (\ref{eq:rrms:propto:wd})
  that if our ball from the example above with \(\theta = 0.5\)
  is ``stretched'' by a factor of 600 to the size of the WD,
  leaving the mass of \(1M\odot\),
  the displacement error will also decrease by a factor of \(\sim 600\)
  and will be only \(\sim\)0.0003\% of the radius of the WD \(\sim 6000\)~km
  at \(\sim 4 \cdot 10^4\;[u_\mathrm{time}]\).
This means that the \phantomsph code can be used
  without corrections for objects like WD, not to mention
  large stars (\ref{eq:rrms:propto:starrel}) or planetary nebulae.
It should be noted, however, that if we consider much larger timescales,
  covering dozens of orbital revolutions, for example in the case of a WD binary system --
  the resulting error in the center of mass displacement also becomes comparable to the radius of a WD.

\begin{figure}[!htb]
  \centering
  \includegraphics[width=0.8\textwidth]{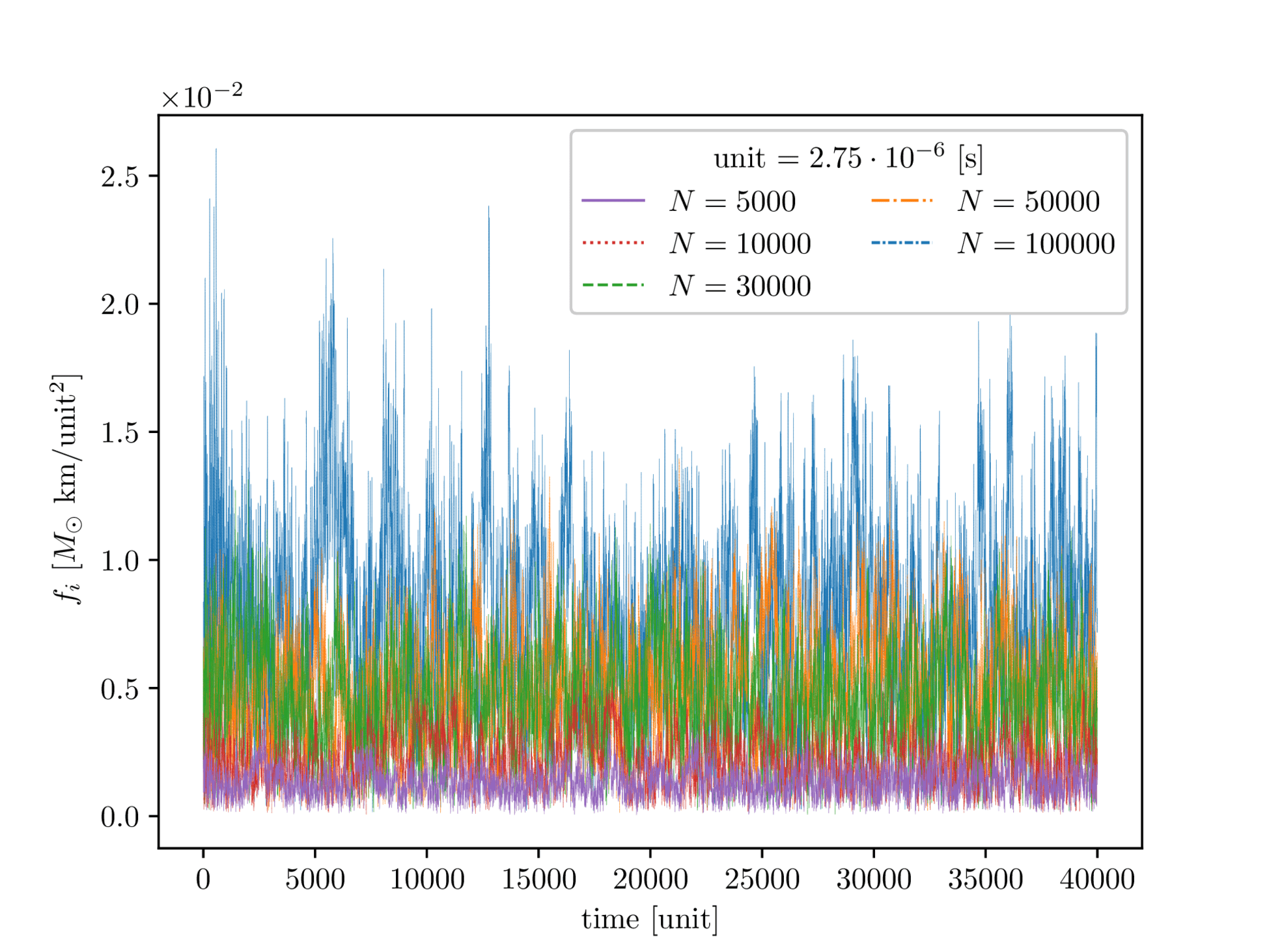}
  \caption{
    The modulus of the uncompensated force
      acting on the center of mass of a single neutron star\\
      calculated for three different numbers
      of SPH-particles at \({\theta = 0.5}\).
  }
  \label{fig:fcom}
\end{figure}

However, from (\ref{eq:rrms:propto:ns}) it follows that
  for \({\sigma > 1/6}\), no changes in \(N\) can eliminate
  the error in the displacement of the center of mass for NS.
To estimate \(\sigma\) we construct the fit \(N^\sigma\)
  of the function \(F(N) = \max \limits_i f_i(T, N)\).
For random variables of modulus of the uncompensated force \(f_i(T, N)\)
  with \(\theta = 0.5\)
  (see Fig.~\ref{fig:fcom}) the value of \(\sigma\) is about \(0.55\).
The same fitting for different values of
  tree opening parameter \(\theta\) shows
  that \(\sigma\) is a monotonically increasing function of \(\theta\).
At \(\theta = 0.2\) this function takes the value \(\sigma \approx 0.15\),
  which is slightly less than \(1/6 \approx 0.16\).

Thus, for values of the tree opening parameter \(\theta \gtrsim 0.2\),
  the standard deviation \(\sqrt{\langle r^2 \rangle}\) will grow with increasing \(N\),
  while for small \(\theta \lesssim 0.2\) it will decrease.
One can always find such a pair \(N\) and \(0 < \theta < 0.2\),
  where the error for~NS becomes negligible.

However, we can see from the formula (\ref{eq:rrms:propto:ns})
  that for any small \(\theta\) the lower bound
  on the standard deviation \(\sqrt{\langle r^2 \rangle}\)
  is a very slow function~\(N^{-1/6}\).
Thus, to reduce the error by one order the number of SPH-particles
  should be increased by 6 orders of magnitude.
In its turn, for any fixed \(N\), the decrease of \(\theta\) leads to an increase of the radius
  \(r \sim 1/\theta\) of the short-range interaction region
  and to the dominance of \(a_\mathrm{short}\)
  defined in (\ref{eq:acceleration}).
This leads to a significant rise
  in computation time (see~Fig.~\ref{fig:rcom_log_theta0-0_5}).
This means that in order to use \phantomsph
  in calculations for objects like NS,
  it is necessary to correct the system for non-conservativity problems
  we demonstrated above.

\section{Conclusion}
\label{sec:conclusion}

In this article, we discussed a way to implement the FMM
  in the \phantomsph code for calculation of the self-gravity forces.
The code is widely used
  in various fields of astrophysics
  \cite{2018ApJ...860L..13P,
    2019ApJ...872..163G,
    2020A&A...641A..64H,
    BlinnikovEtAl2022,
    PotashovYudin2023,
    YudinEtAl2023}.

The standard implementation of the FMM
  (see, e.g., \cite{Dehnen2000, Dehnen2002})
  implies the second order Taylor expansion
  of the gravitational interaction force
  between two cells of the kd-tree
  by small displacements
  of the source-particle and the sink-particle
  expressed through the vector \(\boldsymbol{r}\) connecting the centers of these cells.
Thus, the maximum degree of \(1/r\) will be 4.
In contrast, in \phantomsph, such expansion is carried out step-by-step.
The initial step is to perform the second order Taylor expansion
  for the small displacements of the source-particle.
Afterwards, the second order Taylor expansion
  is carried out for the small displacements of the sink-particle.
The maximum degree of \(1/r\) in this case will be 6.
In this paper, it was shown that even with this
  method of recording the force for any pair of kd-tree cells,
  in the case of mutual interaction
  the Newton's third law is satisfied.
However, linear momentum is not conserved for the entire system.
This fact is explained by that only pairs of kd-tree cells
  ``leaf cell \(\leftarrow\) super-cell'' are considered in \phantomsph.
Reverse pairs are not considered.
In contrast, the method outlined in \cite{Dehnen2000, Dehnen2002}
  involves recursive tree-walk,
  whereby all possible symmetric pairs of well-separated cells
  of all types are considered,
  in accordance with the symmetric opening criterion.

It was shown in the paper that an additional non-physical force,
  resulting from the non-conservation of linear momentum,
  causes the system as a whole to migrate.
The law of this migration is described
  by random memory walk.
Using the example of a system describing a NS,
  it was demonstrated how the magnitude of the mean-square displacement
  of the center of mass of the star from
  its initial position depends on the mass of the star,
  its radius, and the number of SPH-particles in it.
In the case of using \phantomsph for hydrodynamic modelling
  of objects with NS characteristics,
  the shift can indeed be significant.
Thus, for a pair of NS,
  the displacement of the center of mass
  is comparable to the radius of NS
  at the time of a few tens
  Keplerian revolutions of the pair.
For nebulae, hot stars
  and even WD
  such a shift can be
  considered to be negligibly small
  over the same duration.

It was explained that increasing the number of SPH-particles
  does not lead to a decrease of this displacement
  at values of the tree opening parameter \(\theta \gtrsim 0.2\).
The displacement error reduces as the tree opening parameter
  \(\theta \lesssim 0.2\) become smaller,
  but this leads to a significant growth in computation time.
This means that the \phantomsph requires
  a correction using the method
  \cite{Dehnen2000, Dehnen2002}.
However, it is well known that the FMM method,
  even in its realisation \cite{Dehnen2000, Dehnen2002},
  is inherent in the non-conservation of the angular momentum of the system \cite{Marcello2017}.
This aspect requires further investigation.
For conservativity in both linear
  and angular momentum
  the current implementation of \phantomsph
  requires corrections.

The authors would like to thank S.I.~Blinnikov
  for useful discussions and important comments
  and to O.G.~Olkhovskaya for assistance in translation.
The authors are grateful to the anonymous referee for valuable comments.


\sloppy
\printbibliography

\end{document}